\documentclass{article}
\usepackage{graphicx,amsmath,amsfonts,amssymb,amsxtra}
  \usepackage{caption}
\usepackage{txfonts}

\newcommand{\A}{$\text{\AA}$}
\newcommand{\grad}{\ensuremath{^\circ}}
\newcommand{\tn }{T$_\text{N}$}

\newcommand{\lno}{LaNiO$_{3}$}

\newcommand{\rno}{\textit{R}NiO$_{3}$}
\newcommand{\lmo}{La\textit{M}O$_{3}$}

\newcommand{\tmi}{T$_{\mathrm{MI}}$}

\title{Antiferromagnetic correlations in the metallic strongly correlated transition metal oxide LaNiO$_3$}

\author{H. Guo$^{1,*}$, Z. W. Li$^{1,*}$, L. Zhao$^{1}$, Z. Hu$^{1}$, C. F. Chang$^{1}$, C.-Y. Kuo$^{1}$,\\ 
W. Schmidt$^{2}$, A. Piovano$^{3}$, T. W. Pi$^{4}$, O. Sobolev$^{5}$, D. I. Khomskii$^{6}$, L. H. Tjeng$^{1}$\\
 \& A.~C.~Komarek$^{1,\dag}$\\
* These authors contributed equally to this work\\
$\dag$: Alexander.Komarek@cpfs.mpg.de}

\begin{document}

\maketitle

\begin{itemize}
\item $^1$ Max-Planck-Institute for Chemical Physics of Solids, N\"{o}thnitzer Str. 40, Dresden D-01187, Germany
\item $^2$ Forschungszentrum J\"{u}lich GmbH, J\"{u}lich Centre for Neutron Science at ILL, CS 20156, 71 avenue de Martyrs, 38042 Grenoble, France
\item $^3$ Institut Laue-Langevin (ILL), 71 avenue des Martyrs, F-38042 Grenoble Cedex 9, France
\item $^4$ National Synchrotron Radiation Research Center (NSRRC), 101 Hsin-Ann Road, Hsinchu 30077, Taiwan
\item $^5$ Georg-August-Universit\"{a}t G\"{o}ttingen, Institut f\"{u}r Physikalische Chemie, Tammannstrasse 6, D-37077 G\"{o}ttingen, Germany
\item $^6$ Physics Institute II, University of Cologne, Z\"{u}lpicher Str. 77, 50937 Cologne, Germany
\end{itemize}

\section*{Abstract   }
The material class of rare earth nickelates with high Ni$^{3+}$ oxidation state is generating
continued interest due to the occurrence of a metal-insulator transition with charge
order and the appearance of non-collinear magnetic phases within this insulating regime.
The recent theoretical prediction for superconductivity in \lno\  thin films has also triggered
intensive research efforts. \lno\ seems to be the only rare earth nickelate that stays metallic
and paramagnetic down to lowest temperatures.
So far, centimetre-sized impurity-free single crystal growth has not been reported for the rare earth nickelates
material class since elevated oxygen pressures are required for their synthesis.
Here, we report on the successful growth of centimetre-sized \lno\ single crystals by the
floating zone technique at oxygen pressures of up to 150 bar.
Our crystals are essentially free from Ni$^{2+}$ impurities and
exhibit metallic properties together with an unexpected but clear
antiferromagnetic transition.

\section*{Introduction}
The rare earth nickelates \rno\ (\textit{R}~$=$~rare earth, Y) with the high Ni$^{3+}$
oxidation state have continued to attract enormous interest due to the famous bandwidth
controlled metal-insulator (MI) transition and associated unusual charge and spin order
phenomena occurring in this system \cite{Torrance1992,Medarde1997} with even the
possibility for multiferroicity \cite{Giovanetti2009}.
The description of the underlying physics of these phenomena turned out to be a true intellectual
challenge and it is not a surprise that novel theoretical concepts have been (and still
need to be) developed along the way
\cite{Mizokawa1995,Mizokawa2000,Mazin2007,Lee2011,Park2012,Johnston2014,Jaramillo2014,
Subedi2015,Green2016}.
More recently, the prediction of high-T$_{\mathrm{C}}$ superconductivity in \lno\ based
heterostructures \cite{Chaloupka2008} has triggered a flurry of new activities on \lno-\lmo\
superlattices (\textit{M}~=~other metal ion) \cite{Scherwitzl2011,Middey2016}.

Apart from \lno\ all \rno\ (\textit{R}~$=$~Pr-Lu, Y) compounds exhibit a MI transition
at lower temperatures \cite{Torrance1992,Medarde1997} with an insulating
antiferromagnetic ground state. With decreasing \textit{R}-ionic radius the octahedral
tilts become larger. Thus, the Ni-O-Ni bond angles become smaller which alters the
electronic bandwidth and the magnetic exchange interactions. Whereas an enhancement
of the insulating properties and an increase of the MI transition temperature \tmi\ can
be observed for decreasing \textit{R}-ionic radius, the antiferromagnetic transition
temperature and the strength of the exchange interactions decrease with decreasing
\textit{R}-ionic radius and Ni-O-Ni bond angles. With La having the largest ionic radius
of the series one might expect the strongest antiferromagnetic properties for \lno.
However, so far it was reported that \lno\ does not show any magnetic order and, thus,
violates the trend. It remains a paramagnetic metal down to lowest temperatures with an enhanced
effective mass \cite{Torrance1992,Medarde1997,Xu1993,Zhou2014}.

Here, we report on the successful growth of large \lno\ single crystals by the floating zone
technique. Electrical resistivity and Hall effect measurements on our single crystals show
that \lno\ is intrinsically not a bad metal as recently discussed for the \rno\
(\textit{R}~$=$~Pr-Lu, Y) compounds in their paramagnetic phase \cite{Jaramillo2014}.
On the contrary, we found that \lno\ has a high conductivity. Moreover,
we were able to observe bulk antiferromagnetism in magnetization, specific heat and
neutron scattering experiments. Thus, \lno\ appears to be a highly metallic and
antiferromagnetic transition metal oxide - a rather rare combination in oxides.
Special about \lno\ is that it is also very close to an insulating state, making \lno\ an
intriguing quantum material, probably close to a quantum critical point, where strong local electronic correlations at the Ni sites
are likely to interfere in an intricate manner with Fermi surface effects.

\section*{Results}

\subsection*{Crystal growth and characterization}
Using high oxygen pressures of 130-150~bar we were able to grow
\lno\ single crystals with the floating zone technique, see inset of
Fig.~1(a). Further details can be found in the Methods Section and
in the Supplementary Notes~1~\&~2, in the Supplementary Figures 1~\&~2 and in the Supplementary
Movie~1. Powdered crystals
exhibit no impurity phases within the accuracy of powder X-ray
diffraction measurements, see Fig.~1(a). The single crystallinity of
our \lno\ crystal is confirmed by a series of Laue diffraction
patterns from different directions and across the length of this
crystal as well as by single crystal X-ray diffraction, see inset of Fig.~1(a) and Supplementary Figure~1
as well as Supplementary Note~3, Supplementary Figure~3 and Supplementary Table~1. 
Due to the occurrence of a cubic to rhombohedral phase transition
(\textit{Pm$\mathit{\overline{3}}$m}~$\rightarrow$~\textit{R$\mathit{\overline{3}}$c})
somewhat below 1100~K \cite{Medarde1997} our floating zone grown
single crystals are twinned.
 
Fig.~1(b) shows the Ni-L$_{2,3}$ X-ray absorption spectroscopy (XAS)
data of the \lno\ single crystal taken at 300~K and 90~K together
with that of a NiO single crystal serving as a Ni$^{2+}$ reference compound.
We have removed the La-M$_4$ white lines located at 850.6~eV from
the \lno\ spectra using the La-M$_4$ spectrum of LaCoO$_3$. It is
well known that XAS spectra at the L$_{2,3}$ edge of transition
metal oxides are highly sensitive to the valence state, in
particular, an increase of the valence state of the transition metal
ion by one causes a shift of the XAS L$_{2,3}$ spectra by one or
more eV towards higher energies \cite{xasA}. The more than one eV
higher energy shift between the spectra of NiO and LaNiO$_3$
indicates the formal Ni$^{3+}$ valence state in \lno\ \cite{xasA,xasC},
with the note that the spectrum can not be interpreted in terms of a Ni 3d$^{7}$ configuration,
but, rather by a coherent mixture of 3d$^{8}$ and 3d$^{8}$$\underline{L}$$\underline{L}$ configurations \cite{Mizokawa1995,Mizokawa2000,Johnston2014,Green2016},
where each $\underline{L}$ denotes a hole in the oxygen ligand.
Here we can exclude Ni$^{2+}$ impurities in our \lno\ single crystal - otherwise the sharp
main peak of the Ni$^{2+}$ impurity spectrum would have been visible as a sharp shoulder at the
leading edge \cite{xasA}. At the O-K edge, the pre-edge peak is shifted by about one eV to
higher energies when going from Ni$^{3+}$ to Ni$^{2+}$ \cite{xasA,xasC}. Fig.~1(c) shows
the O-K XAS spectra of our \lno\ single crystal (blue) and the NiO reference compound (red).
The O-K XAS spectra demonstrate even more clearly that there is no spectral feature from
Ni$^{2+}$ impurities and thus that our as-grown \lno\ single crystals are highly stoichiometric.
This is further confirmed by thermogravimetric and inductively coupled plasma optical emission spectroscopy
measurements, see Methods section and the Supplementary Note~2.

\subsection*{Temperature dependence of physical properties}

In Fig.~2(a) we show the temperature dependence of the lattice
parameters of our \lno\ single crystal that has been powdered and
measured by means of powder X-ray diffraction. There is no clear
indication for the presence of a structural anomaly which otherwise occurs
readily in the other nickelates \rno\ with smaller rare earths (\textit{R}~$=$~Pr-Lu, Y) when
cooling through the metal-insulator transition
\cite{Torrance1992,Medarde1997}.

The electrical resistivity $\rho$ of our \lno\ single crystal that is shown in Fig.~2(b)
is in the $\muup\Omega$~cm range, reaching $\sim$6~$\muup\Omega$~cm at low temperatures, i.e.
distinctly lower than that of \lno\ powder samples \cite{Torrance1992,Medarde1997,Xu1993,Zhou2014}
or \lno\ thin films \cite{Scherwitzl2011} or of crystals grown under 50~bar oxygen pressure \cite{mitchell2017}. 
Thus, our \lno\ single crystals are more metallic, see also Supplementary Note~4 and Supplementary Figure~4(a). 
Also the measurement of the Hall effect on the single crystal \lno\ (for temperatures between 2~K and 300~K) yields high carrier densities of
about 4.8$\times$10$^{28}$~m$^{-3}$, corresponding to $\sim$2.7 (close to 3) electrons per
formula unit, consistent with the 3+ valence of the Ni ions in \lno.
Up to $\sim$35~K the low-temperature behavior of the resistivity is Fermi-liquid-like,
$\rho$($T$)~=~$\rho_0 + A T^{n}$, with $\rho_0$~$\sim$~6.45~$\muup\Omega$~cm,
$A$~=~1.62$\cdot$10$^{-3}$ $\muup\Omega$~cm~K$^{-2}$ and $n$~$\sim$~2.0, see Fig.~2(b).
The value of $A$ is of the same order as reported in literature~\cite{zhou2005}.
A fit over the entire temperature range gives an exponent $n$~$\sim$~1.50(1)
with the residual resistivity $\rho_0$~$\sim$~4.83$\muup\Omega$~cm, similar to the value of the fit at low temperatures.
Most probably this behavior is a signature that \lno\ is close to a quantum critical point
(cf. the results for PrNiO$_3$ under pressure \cite{zhou2005}).

Fig.~2(c) displays the temperature dependence of the magnetic susceptibility $\chi$ of our \lno\
single crystal, see also Supplementary Note~4 and Supplementary Figure~4(b). First of all, we notice that it shows a significantly smaller low-temperature
upturn than reported previously for powder and ceramic samples \cite{Xu1993,Zhou2014}, which
confirms unprecedented high quality of our single crystals. Surprisingly, $\chi$ exhibits an anomalous kink at
$\sim$157~K which we take as an indication for a hitherto unknown antiferromagnetic transition
in \lno. That this anomaly is not simply caused by a signal from a tiny fraction of a magnetic impurity
phase (which is so small that it is not visible in our powder X-ray diffraction measurements) can
be excluded by our specific heat ($C_p$) measurements. As can be seen in Fig.~2(d) there is a
small but clearly visible anomalous peak at $\sim$157~K in $C_p/T$. Also the resistivity data excludes
that the anomaly in the susceptibility is caused by the presence of oxygen deficient \lno\
minority phase that becomes antiferromagnetic and insulating at low temperatures
\cite{Sanchez1996,Gayathri1998}: we do not observe an upturn or a slowing down of the decrease in the resistivity on cooling. 
Moreover, our samples have resistivities in the $\muup\Omega$~cm range and have conductivities higher than reported so far.
All these support the notion that the transition is an intrinsic bulk property and not due to an impurity phase.  The
low-temperature behaviour of the specific heat supports a Fermi-liquid type of the ground state
of \lno, $C_p$($T$)~=~$\gamma T + \beta T^3$,
with $\gamma$~$\sim$~17~mJ mol$^{-1}$ K$^{-2}$, consistent with the value 18~mJ mol$^{-1}$ K$^{-2}$ reported before \cite{Zhou2014},
and with $\beta$~=~1.87(5)$\cdot10^{-4}$~J~mol$^{-1}$K$^{-4}$, thus yielding a Debye temperature $\theta_D = (12\pi^4NR/5\beta)^{1/3}$ = 373 K, where $N$ is the number of atoms in the chemical formula and $R$ is the ideal gas constant.

\subsection*{Neutron scattering experiments}
The availability of sizeable \lno\ single crystals also enabled us to study this intriguing system by
means of neutron diffraction and inelastic neutron scattering. These experiments were performed
at the Thales, IN12 and IN8 spectrometers at the ILL in Grenoble, France. 
Within elastic scans we were able to observe quarter-integer peaks at low temperatures, see Fig.~3(a). These
quarter-integer peaks (in pseudocubic notation) resemble those found in powder neutron
diffraction experiments within the insulating regime of \rno\ \cite{pwdn}. The propagation vector
observed for the insulating antiferromagnetic regime of \rno\ amounts to (1/4~1/4~1/4) in
pseudocubic notation (or (1/2~0~1/2) in orthorhombic notation) \cite{Medarde1997}. 
Moreover, the study of the temperature dependence of these quarter-integer peaks 
indicates an onset temperature which coincides with the magnetic ordering temperature \tn\ that we 
observed in susceptibility and specific heat measurements of \lno, see Fig.~3(b).

The magnetic origin of these quarter-integer peaks in \lno\ could be also confirmed by polarized neutrons at the
IN12 spectrometer. In Fig.~4(a,b) $L$-scans across two quarter-integer peaks are shown for
the three spin flip (SF) channels and for a non-spin-flip channel. Only in the spin flip channels
neutron scattering intensities $\sigma_{x_i x_i}$ can be detected. This unambiguously shows
the magnetic nature of these quarter-integer reflections in \lno. 
Although we can not detect a symmetry lowering from our high resolution powder X-ray diffraction measurements
 - see Fig.~2(a) - the pseudocubic propagation vector for rhombohedral \lno\ is the same as the pseudocubic propagation
vector for the orthorhombic insulating nickelates \rno\ ($R$~=~Pr-Lu, Y).
Based on the magnetic symmetry analysis for the high symmetry cubic structure with space group \textit{Pm$\mathit{\overline{3}}$m} and for the propagation vector (1/4 1/4 1/4)
a helical magnetic structure with moments spiraling perpendicular to the propagation vector is consistent with our single crystal neutron data, see Supplementary Note~5, the
Supplementary Tables 2~\&~3 and the Supplementary Figure 6.
The size of the ordered moment amounts to $\sim$0.3~$\muup_\mathrm{B}$ which indicates that magnetism in \lno\ is a bulk property and does not originate from a tiny (insulating) impurity phase. This small magnetic moment might explain why paramagnetic properties have been reported for \lno\
in the past \cite{Torrance1992,Medarde1997,Xu1993,Zhou2014}.

We also have been able to study and observe the magnetic excitations in \lno\ by means of inelastic neutron scattering
thereby also providing further support that the antiferromagnetism is a bulk property.
As can be seen in Fig.~5, magnetic excitations are clearly visible up to at least 12~meV.
With increasing energy transfer the magnetic peaks become somewhat broader and
damped which is indicative for the appearance of fluctuations in this itinerant
antiferromagnetic system. These fluctuations could be also responsible for a reduced
ordered moment of $\sim$0.3~$\muup_\mathrm{B}$ in \lno. 

\section*{Discussion}

\lno\ appears to be a rare case of an antiferromagnetic and metallic transition metal
oxide with a fully 3D crystal and electronic structure. Other systems like (La,Sr)$_3$Mn$_2$O$_7$
and Ca$_3$Ru$_2$O$_7$ have a lower electronic and structural dimensionality where the
antiferromagnetic order is resulting from a stacking of ferromagnetic layers
\cite{manganite,ruthenate}. So far, the chromate system $A$CrO$_3$ ($A$~=~Ca, Sr) and
very recently, RuO$_2$ are known to be the only other intrinsically antiferromagnetic and
metallic transition metal oxides with such a fully 3D crystal and electronic structure
\cite{cacr,Berlijn2017}. The systems CaCrO$_3$ and \lno\ have in common that the oxidation
state of the transition metal ion is very high. Thus, the oxygen $2p$  to transition metal $3d$
charge transfer energy here is apparently negative \cite{Mizokawa2000,Green2016,KhomskiiBook}
resulting in extreme $2p$-$3d$ covalency where the presence of holes in the oxygen band can
effectively prevent the opening of the conductivity gap and at the same time mediate strongly
the magnetic exchange interactions. However, in contrast to CaCrO$_3$ \cite{cacr}
the rare earth nickelate \lno\ in single crystalline form is much more metallic and shows
conductivities in the $\muup\Omega$~cm range. This is probably also true for RuO$_2$ \cite{Berlijn2017}.
Unique for \lno\ is that it is close to the insulating phase of \rno\ (\textit{R}~$=$~Pr-Lu, Y) indicating
the importance of strong correlation effects, making it rather exceptional among all transition metal oxides.

Although charge order - which plays a very important role in nickelates \rno\ with small rare earth ions
\textit{R} - cannot be observed in our present XRD and neutron diffraction measurements - see Fig.~4(c) -
the presumably associated symmetry lowering of the structure that goes along with the antiferromagnetic
order may produce only very weak new peaks in diffraction experiments. Some hints for this may in fact
be found in a recent pair distribution function study using neutron diffraction \cite{Li2016}.
Nevertheless, with the charge order effects being so weak, one could infer that \lno\ is perhaps better
described using Fermi surface arguments \cite{Lee2011,Berlijn2017} while the other insulating nickelates \rno\
with the smaller $R$-ionic sizes and with higher charge ordering temperatures can be more intuitively
understood in terms of local charge or bond disproportionations
\cite{ Mizokawa2000,Mazin2007,Park2012,Johnston2014,Subedi2015,Green2016}.
Note, that recent ab-initio calculations \cite{SubediARX} reproduced metallic and antiferromagnetic state of \lno,
and also predicted charge (or rather bond) disproportionation, whose magnitude is however below our experimental detection limit. 
One needs special dedicated experiments to probe for the eventual symmetry lowering with the appearance of inequivalent Ni ions.
However, our main conclusion - the existence of antiferromagnetic ordering in highly metallic single crystals of \lno,
is confirmed by these calculations.
According to our findings we now also present a tentative \rno\ phase diagram in Fig.~6.

\section*{Methods}
\subsection*{Chemical synthesis}
The \lno\ single crystal was grown under high oxygen pressures of
130-150~bar with a growth speed of 6-7~mm h$^{-1}$ in a one-mirror furnace
from \textit{Scidre} that was equipped with a 5000~W Xe-lamp and with
counter-rotation of feeding and seeding rods - see Supplementary
Movie 1.
The seeding and feeding rods were synthesized by pressing appropriate
mixtures of La$_2$O$_3$ and NiO into rods with roughly 10~cm length and 6~mm diameter.
These rods were sintered at 800\grad C-1000\grad C for several days.
The temperature of the melting zone was measured in-situ by means of a pyrometer and is close to 1800\grad C at $\sim$140 bar p$_{\mathrm{O_2}}$.

\subsection*{X-ray diffraction}
Laue diffraction measurements were performed on a multiwire real-time
back-reflection Laue camera from \textit{Multiwire Laboratories}; see also the
Supplementary Note~1 and Supplementary Figure~1.

Powder X-ray diffraction measurements were performed on a
\textit{Bruker D8 Discover A25} diffractometer which is equipped with
a Johansson monochromator for Cu K$_{\alpha_1}$ radiation. A closed
cycle helium cryostat (\textit{Phenix} of \textit{Oxford Cryosystems})
was used for temperature-dependent measurements.

Single crystal X-ray diffraction measurements have been performed on
a twined single crystal of LaNiO$_3$ 
using a \textit{Bruker D8 VENTURE} single crystal X-ray diffractometer equipped with a bent graphite monochromator for Mo $K_{\alpha}$ radiation (about 3$\times$ intensity enhancement) and a \textit{Photon} CMOS large area detector.
A crystal with roughly 20~$\muup$m diameter has been measured and a multi-scan absorption correction has been applied to the data (minimum and maximum transmission:
0.6184 and 0.7519 respectively).
12328 (observed) reflections (H: -12~$\rightarrow$~13, K: -13~$\rightarrow$~9 and L: -32~$\rightarrow$~31) have been collected with an internal R-value of 6.70\%, a redundancy of 36.8 and with 98.85\%\ coverage up to 2$\Theta_{max}$~=~123.6\grad.
For the refinement the \textit{Jana2006} program package was used.
The Goodness of fit of our crystal structure refinement amounts to 1.93 and the R- and weighted R-values amount to   2.29\%\   and  5.79\%\ respectively.
The refinement of the Ni and La occupancies yields an almost stoichiometric composition: La$_{0.995(12)}$Ni$_{1.000(14)}$O$_3$.
The structural parameters are listed in Supplementary Table~1 and the crystal structure is visualized in the Supplementary Figure~3. 

\subsection*{Composition determined by ICP and TG measurements}
Inductively coupled plasma optical emission spectroscopy (ICP-OES) measurements yields the following composition of \lno:
La$_{1.001(11)}$Ni$_{0.999(4)}$O$_{3+\delta}$. 
Moreover, thermogravimetric measurements confirm an almost perfectly stoichiometric oxygen content with $\delta = -0.002$.

\subsection*{X-ray absorption spectroscopy}
XAS measurements have been performed at the 08B beamline of the
National Synchrotron Radiation Research Center (NSRRC), Taiwan.

\subsection*{Magnetization measurements}
The magnetic properties were studied using a \textit{Quantum Design Inc.
MPMS-5XL} SQUID magnetometer.

\subsection*{Electrical conductivity}
The measurements of electrical resistivity and specific heat were
carried out using a four-probe and a standard thermal relaxation
calorimetric method in a \textit{Quantum Design Inc.} Physical
Property Measurement System (\textit{PPMS}).

\subsection*{Neutron measurements}
Unpolarized neutron measurements have been performed at the IN8 and
Thales Spectrometers at the ILL in Grenoble, France. 
For the elastic and inelastic measurements on the Thales spectrometer a pyrolytic graphite (PG) (002)
monochromator and analyzer as well as a velocity selector were used
($k_\mathrm{f}$~=~1.8 \A$^{-1}$). On the IN8 spectrometer very first quick
inelastic scans have been made using a Si monochromator and a
Flatcone analyzer array. Polarized neutron measurements have been
performed at the IN12 Spectrometer at the ILL in Grenoble, France. A
velocity selector was used for choosing the incident neutron
wavevector $k_\mathrm{i}$ of 2.25~\A$^{-1}$. A horizontally and vertically
focussing PG (002) monochromator and Heusler
(111) analyser have been used. The incident neutron beam was
polarized by a transmission polarizer (cavity) in the neutron guide and the measured flipping ratio amounts to
22.2.

\subsection*{Data availability}
The authors declare that the data supporting the findings of this study are 
available within the paper and its supplementary information files. \\

\subsection*{Acknowledgement} 
We thank O. Stockert, G. A. Sawatzky and A. Fujimori for valuable discussions.
We thank M. Schmidt and his team for thermogravimeteric measurements.
We thank G. Auffermann and her team for ICP measurements.
We acknowledge A. Todorova and G. Ryu for support on preparatory work.
We thank the team of H.~Borrmann for initial room-temperature powder X-ray
diffraction measurements used for phase analysis.
The research in Dresden and Cologne is partially supported by the
the Deutsche Forschungs-Gemeinschaft through FOR 1346. \\

\subsection*{Competing Interests}   
The authors declare that they have no
competing financial interests. \\
\subsection*{Author Contributions}  Conceiving experiments and project management: A.~C.~K.;
physical experiments:  H. G., Z. W. L., L. Z., Z. H., C. F. C., C. Y. K., A. T., W. S., A. P. T. W. P., O. S. and A.~C.~K.;
chemical synthesis: A. C. K., H. G.;
interpretation and manuscript writing: A.~C.~K., D. I. K. and L. H. T. \\
H. G. and Z. W. L. are equally contributing authors. \\
\subsection*{Correspondence}
Correspondence and requests for materials
should be addressed to A.C.K.~(email: Alexander.Komarek@cpfs.mpg.de). \\

 \newpage

\begin{figure}
\begin{center}
\includegraphics*[width=0.5\columnwidth]{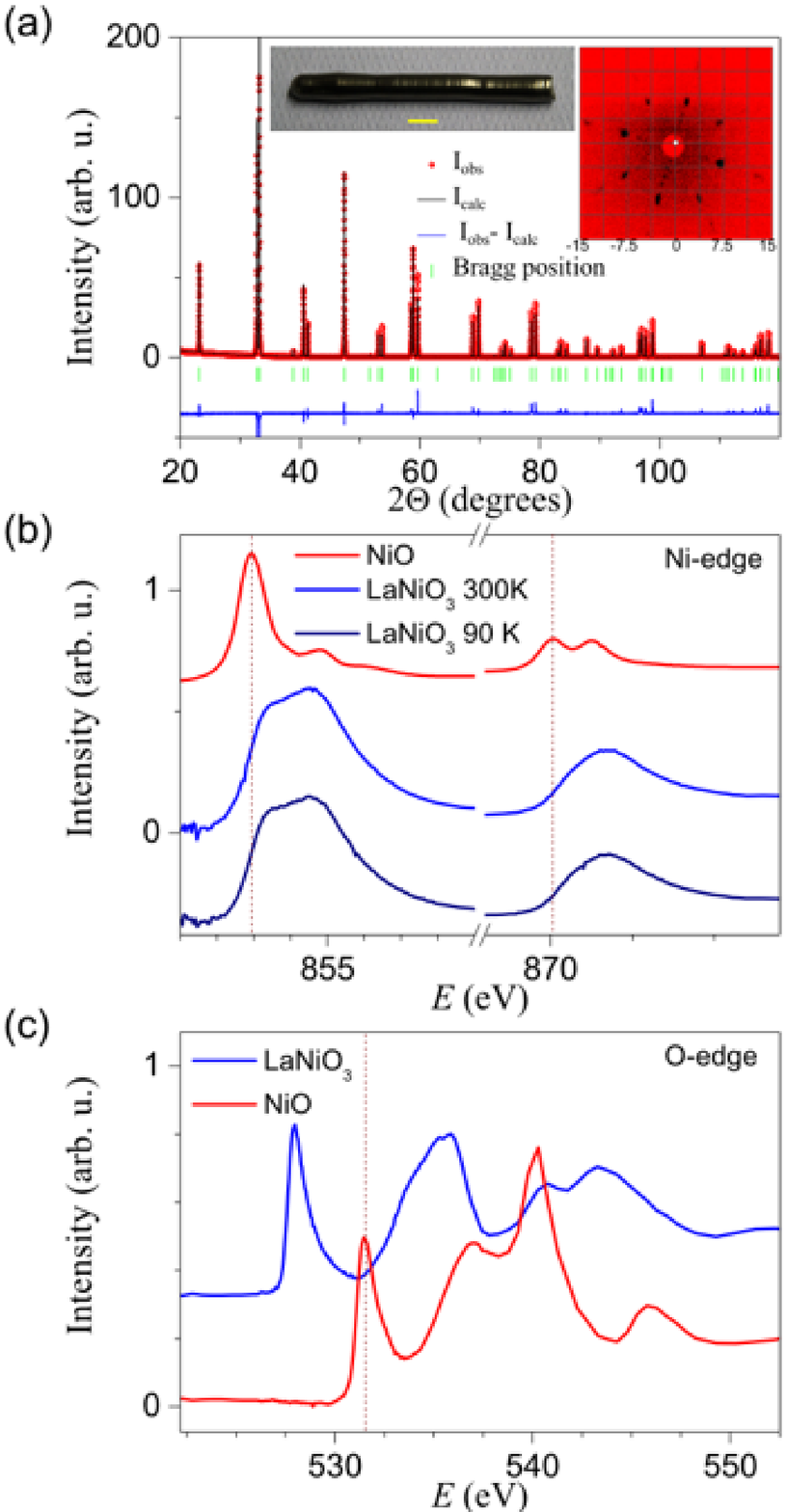}
\end{center}
\caption{\textbf{Characterization of LaNiO$_3$} (a) Powder X-ray diffraction pattern of a powdered \lno\ single
crystal. The inset shows a photo of the \lno\ single crystal and its Laue diffraction
pattern. A length scale of 1~cm is indicated by the yellow bar within the photo of the crystal and 
the distance between two lines in the Laue pattern amounts to 3.75~cm. (b) The Ni-L$_{2,3}$ XAS spectra of the \lno\ single crystal measured
at 90~K and 300~K together with the spectrum of a NiO reference compound.
(c) The O-K XAS spectra of our \lno\ single crystal measured at 300~K (blue line) together with the spectra
of the NiO reference compound (red line).}
\label{fig1}
\end{figure}

\begin{figure}
\begin{center}
\includegraphics*[width=0.5\columnwidth]{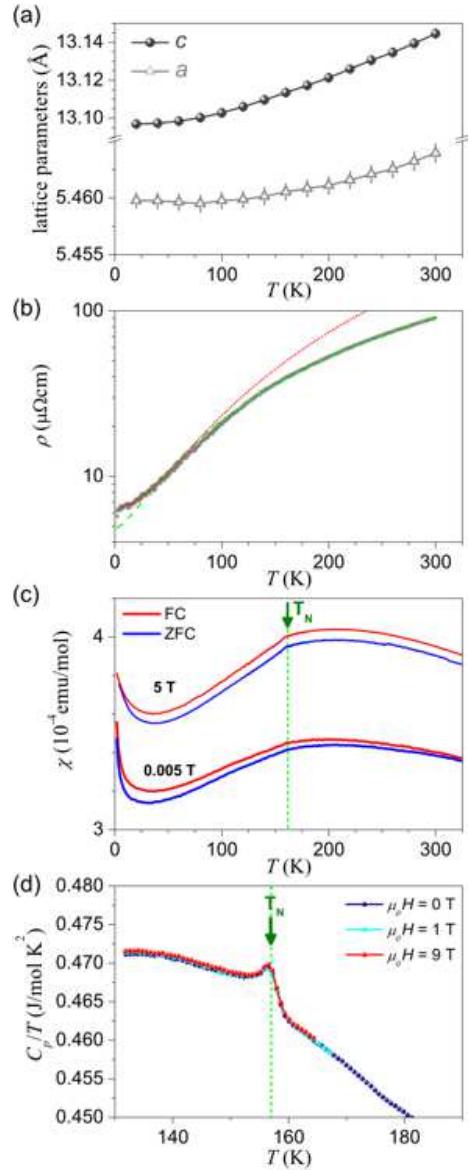}
\end{center}
\caption{\textbf{Temperature dependence of physical properties of LaNiO$_3$} (a) Lattice parameters of our \lno\ as a function of temperature (space group \textit{R$\bar{3}$c}).
The error bars are the standard deviations obtained from Rietveld refinement using the Fullprof software.
(b) Electrical resistivity of \lno. The red line indicates a fit to $\rho_0 + A T^{n}$ (for $T<35$~K) and the dashed green line is a fit over the entire temperature range. (c) Magnetic susceptibility of \lno\ under field cooled (FC) and zero-field cooled (ZFC) conditions for applied fields of $\muup_0$$H$~=~0.005~T and 5~T. The high and low field measurements both together indicate an antiferromagentic transition at \tn\ that is an intrinsic property of \lno. (d) Specific heat measurements of \lno\ in fields of 0~T, 1~T and 9~T. These measurements further corroborate that the antiferromagnetic transition is a bulk property of \lno.}
\label{fig2}
\end{figure}

\begin{figure}
\begin{center}
\includegraphics*[width=0.5\columnwidth]{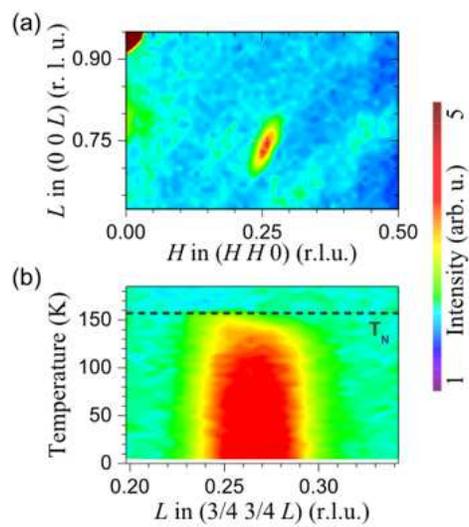}
\end{center}
\caption{\textbf{Single crystal neutron diffraction measurements} (a) Neutron scattering intensities within the (\textit{H H L})
plane of reciprocal space (in pseudocubic notation) of the \lno\ single crystal.
These intensites were obtained at 1.6~K. (b) The temperature dependence of
a quarter-integer peak as measured in (3/4~3/4~$L$)-scans for different temperatures. The dashed line indicates the transition temperature (\tn) observed in the
specific heat. }
\label{fig3}
\end{figure}

\begin{figure}
\begin{center}
\includegraphics*[width=0.5\columnwidth]{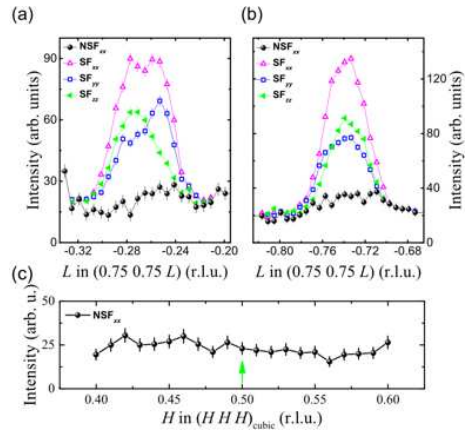}
\end{center}
\caption{\textbf{Polarized neutron diffraction measurements}  (a,b) Polarized neutron scattering experiments for the
\lno\ single crystal at 2~K. Spin-flip (SF) and non-spin-flip (NSF) channels were
measured for neutron spins parallel to the scattering vector (\textit{x}-direction),
perpendicular to the scattering plane (\textit{z}-direction) or perpendicular to
\textit{x} and \textit{z} (\textit{y}-direction).
(c) A structural scan at 2~K showing no indications for any charge ordering peak at the (0.5 0.5 0.5) position in pseudocubic notation (indicated by the green arrow).
 The intensity error bars are statistical error bars calculated by the square root of intensity.}
\label{fig4}
\end{figure}

\begin{figure}
\begin{center}
\includegraphics*[width=0.375\columnwidth]{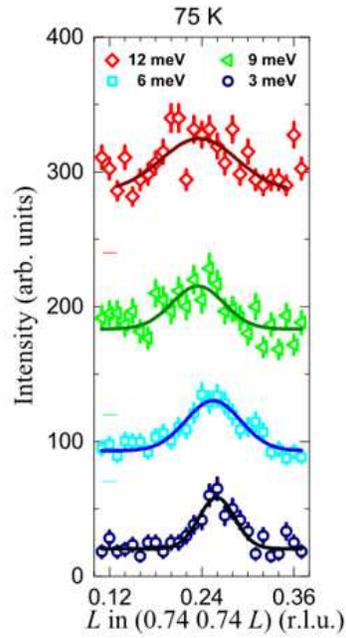}
\end{center}
\caption{\textbf{Magnetic excitations in \lno} \ Inelastic neutron scattering intensities measured  at 75~K for
3, 6, 9 and 12~meV energy transfer. The horizontal bars indicate the shift of the data in vertical direction.
Up to 12~meV the magnetic peaks become broader and damped which is indicative for the appearance of fluctuations 
in this itinerant antiferromagnetic system.
 The intensity error bars are statistical error bars calculated by the square root of intensity.}
\label{fig5}
\end{figure}

\begin{figure}
\begin{center}
\includegraphics*[width=0.6\columnwidth]{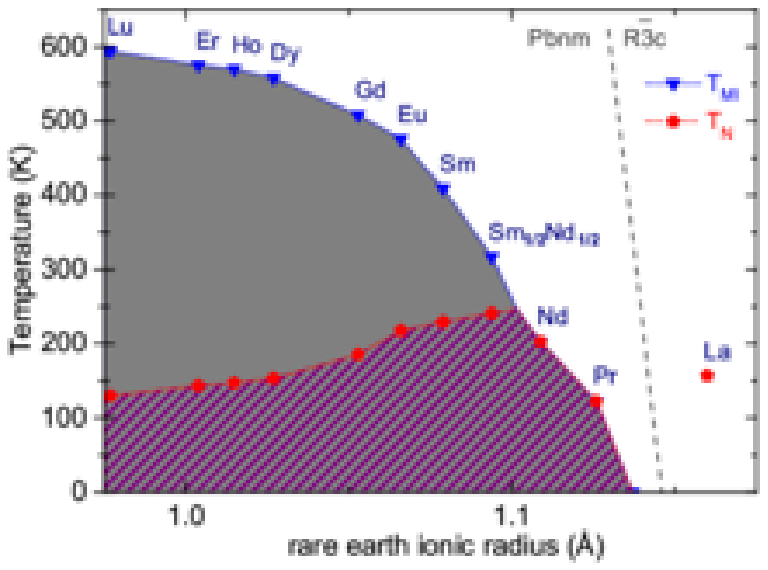}
\end{center}
\caption{\textbf{Tentative \rno\ phase diagram.} The blue triangles indicate the metal-insulator transition temperatures T$_\mathrm{MI}$
associated with the onset of charge ordering and the appearance of monoclinic distortions. The red symbols indicate the N\'{e}el temperatures T$_\mathrm{N}$ associated with the onset of antiferromagnetic ordering. 
Data points for other \textit{R} than La were taken from literature \cite{pwdn,Zhou2003}. The gray dashed curve subdivides orthorhombic (monclinic) and rhombohedral regimes.}
\label{fig6}
\end{figure}

\clearpage
\newpage
\mbox{~}
\clearpage
\newpage

\textbf{{\LARGE Supplementary Information}}

\section*{Supplementary Note 1}
\par\textbf{Laue X-ray diffraction}
\par We have performed Laue diffraction measurements for different spots on one of our LaNiO$_3$ single crystals.
As can be seen in Supplementary Figure~1, for each studied side (left and right side) the diffraction patterns are the same for all  positions (spots) over the entire
length of the crystal.
This indicates, that our crystal is a single crystal.
Also our single crystal has no NiO impurities visible by eyes in powder X-ray diffraction measurements (see Fig.~1 in the main text).

\section*{Supplementary Note 2}
\par\textbf{Thermogravimetric measurements}
\par Supplementary Figure~2 shows a typical thermogravimetric (TG) measurement of one of our nickelate samples 
that was heated to 900$^{\circ}$C in a flow (100ml/min.) of 20\%\ H$_2$ and  80\%\ Ar gas.
The oxygen concentration amounts to $\sim$3.005 according to the graphical analysis and $\sim$2.998 as derived from weighing initial and final masses before and after the reduction of the sample. Both values basically indicate the absence of Ni$^{2+}$ impurities in our LaNiO$_3$ crystals. The small difference between these values is comfortably within the typical error of TG measurements and can be attributed to the impact of the gas flow which is affecting the buoyant force that has non-negligible influence on the weighing of the sample mass during the reaction.
 
\section*{Supplementary Note 3}
\par\textbf{Crystal structure}
\par Moreover, we determined the atomic positions of LaNiO$_3$ by means of single crystal X-ray diffraction.
The structural parameters are listed in Supplementary Table~1 and the (obtained) crystal structure is visualized in Supplementary Figure~3.

\section*{Supplementary Note 4 }
\par\textbf{Physical properties}
\par We also re-measured the magnetic susceptibility and electrical resistivity for six different single crystals from three batches that we have grown, see Supplementary Figure~4.

If the crystal is a single crystal and not polycrystalline it shows the kink in the magnetic susceptibility and a low value of the electrical resistivity, see Supplementary Figure~4.
However, if the crystal is not single crystalline, we also observe a magnetization without any kink, similar as in Ref.~[20].
Therefore, we could imagine that magnetic ordering only appears in LaNiO$_3$ single crystals where domain sizes are very large and samples are very pure.

\par Finally, there seems to be no field dependence (up to 9~T) of the electrical resistivity of \lno\ - see Supplementary Figure~5.
Since the single crystals are twined (and pseudocubic) also no direction dependence on the magnetic field direction can be expected.

\section*{Supplementary Note 5}
\par\textbf{Magnetic structure}

\par Since our single crystal is fully twinned we decided to make a magnetic symmetry analysis for the high symmetry cubic (undistorted perovskite) structure
with space group \textit{Pm$\mathit{\overline{3}}$m}.
The integrated intensities of the (0 0 1) and (1 1 0) nuclear peaks and (0.25 0.25 0.75), (0.75 0.75 0.25) and (0.75 0.75 0.75) magnetic peaks have been obtained from Gaussian fits to the corresponding Q scans (note, that an analyzer was used in our triple-axis measurements,) and normalized by the monitor and by the attenuation factor (plexi: 7.5) for the nuclear reflections. The scale factor was determined from the nuclear intensities and fixed for the subsequent magnetic structure refinement.
The magnetic structure was determined by irreducible representation (IR) analysis for space group $Pm\bar{3}m$ and for a propagation vector (0.25 0.25 0.25). The reducible magnetic representation for the Ni ions at the $1b$ site is decomposed into two IRs as:
 \begin{equation}\label{}
   \Gamma = \Gamma_2 \oplus \Gamma_3.
 \end{equation}
The basis vectors for the IRs are listed in Supplementary Table~2.
and a comparison of observed and calculated intensities is listed in Supplementary Table~3.
For the cubic symmetry the appearance of the (3/4 3/4 3/4) reflection excludes the model with longitudinal modulation
of the magnetic moments (see Supplementary Figure~6(a)) which is also reflected by the large R- and weighted R values that amount to 65.0\%\ and  69.4\%.

The model with helical arrangement of the magnetic moments perpendicular to the propagation vector (see Supplementary Figure~6(b)) is consistent with our data (R- and weighted R values amount to 16.3\%\ and  17.7\%\ respectively) and yields an ordered moment of about 0.3 $\muup_\mathrm{B}$.

\begin{figure}
\includegraphics*[width=0.6\columnwidth]{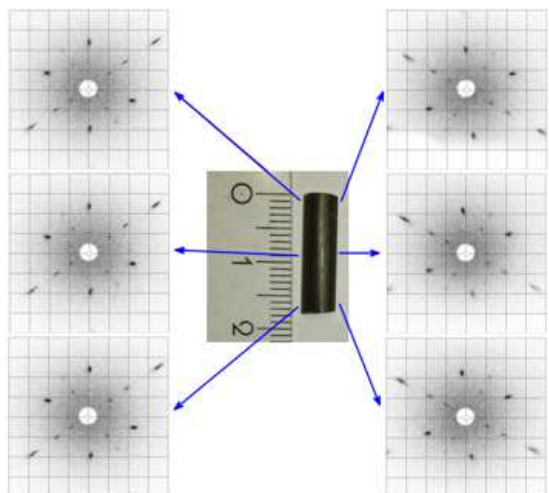}
\caption*{Supplementary Figure 1: \textbf{Laue diffraction images} A photo of one of our LaNiO$_3$ single crystals (batch \# 2) together with Laue diffraction images measured at different spots on the sample which are indicated by the blue arrows.}
\label{fig5}
\end{figure}

\begin{figure}
\begin{center}
\includegraphics*[width=0.6\columnwidth]{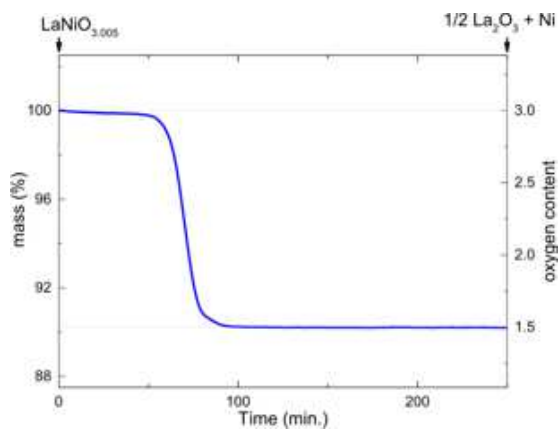}
\end{center}
\caption*{Supplementary Figure 2: \textbf{TG measurements} Temporal dependence of the sample mass of one of our LaNiO$_3$ crystals during the reduction process of a thermogravimetric measurement.
The right scale denotes the calculated oxygen content $x$ as a function of time assuming that the reaction 2~LaNiO$_{x}$~$+$~3~H$_2$~$\rightarrow$~La$_2$O$_3$~$+$~2~Ni~$+$~3~H$_2$O$\uparrow$ went to completion after 250~min. The grey dashed horizontal lines refer to a nominal oxygen content of $x$~=~3 and 1.5 respectively.}
\label{figSpin}
\end{figure}

\begin{figure}
\begin{center}
\includegraphics*[width=0.5\columnwidth]{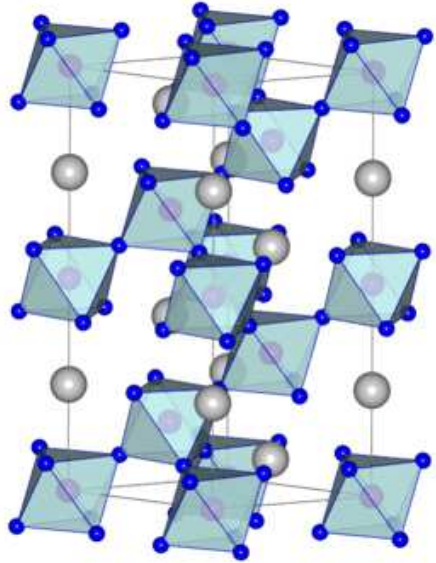}
\end{center}
\caption*{Supplementary Figure~3: \textbf{Crystal structure} A representatiopn of the crystal structure of \lno\ in the hexagonal setting of space group \textit{R$\mathit{\overline{3}}$c} as derived from our single crystal X-ray diffraction measurement at room-temperature.
Magenta/blue/white spheres: Ni, O and La ions.}
\label{fig6}
\end{figure}

\begin{figure}
\includegraphics*[width=0.6\columnwidth]{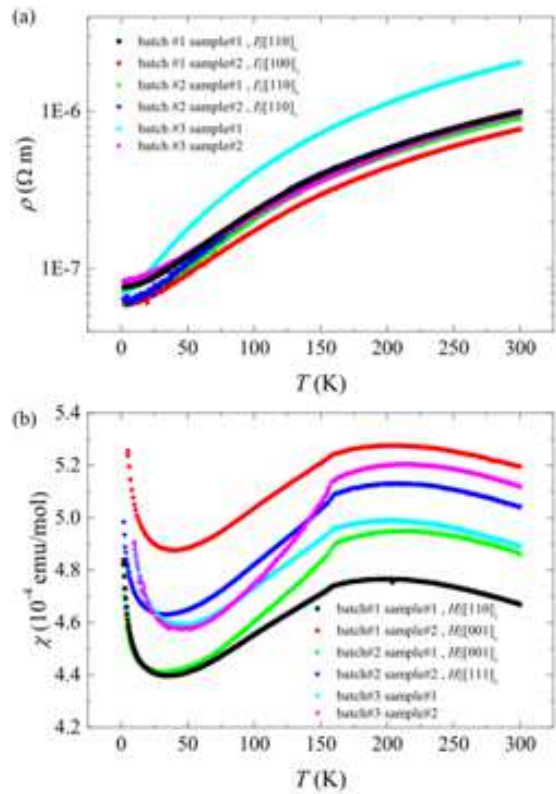}
\caption*{Supplementary Figure 4: \textbf{Physical properties}  (a) Electrical resistivity and (b) magnetic susceptibility measurements which have been done for two pieces (each) of three different single crystals (batches) that we have grown.}
\label{fig6}
\end{figure}

\begin{figure}
\begin{center}
\includegraphics*[width=0.6\columnwidth]{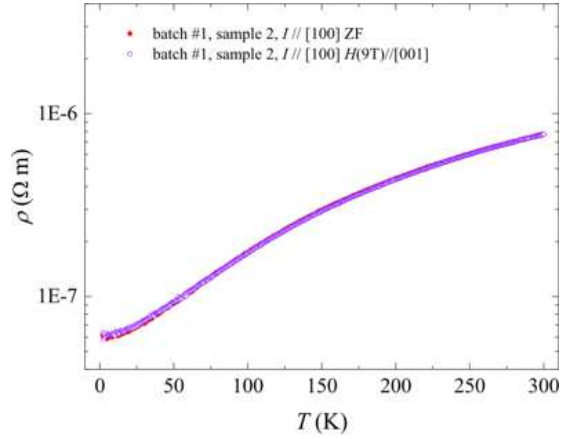}
\end{center}
\caption*{Supplementary Figure 5: \textbf{Field dependence of the electrical resistivity} The electrical resistivity of \lno\ measured for different applied magnetic fields $H$.}
\label{fig6}
\end{figure}

\begin{table}
  \begin{tabular}{l|cccccc}
	\hline
	\hline
    atom & occup. & $x$ & $y$ & $z$ & & \\
   \hline
La1 & 0.995(12) &  0 &  0 &  0.25 \\ 
Ni1 & 1.000(14) &  0 &  0 &  0 \\ 
O1 & 1 &  0.5457(4) &  0 &  0.25 \\ 
   \hline
   \hline
   & & & & & & \\
    atom & U$_{11}$ (\AA$^{2}$) & U$_{22}$ (\AA$^{2}$) & U$_{33}$ (\AA$^{2}$) &  & &  \\
   \hline
   La1 &  0.00523(7) &  0.00523(7) &  0.00803(16)    &  & &  \\
   Ni1 &  0.00304(13) &  0.00304(13) &  0.0055(3)    &  & &  \\
   O1  &  0.0077(5) &  0.0077(9) &  0.0093(11)    &  & &  \\
   \hline
     atom &   U$_{12}$ (\AA$^{2}$) & U$_{13}$ (\AA$^{2}$) & U$_{23}$ (\AA$^{2}$) & U$_{iso}$ (\AA$^{2}$)  &   & \\
   \hline 
   La1 &   0.00261(4) &   0 &   0 &   0.00616(7)   & \\
   Ni1 &  0.00152(7) &   0 &   0 &   0.00388(13)   & \\
   O1 &   0.0038(5) &  -0.0015(4) &  -0.0029(7)   &   0.0082(6)     & \\ 
   \hline
   \hline
  \end{tabular} 
  \caption*{Supplementary Table 1: \textbf{Crystal structure} Refinement results of single crystal X-ray diffraction measurements of \lno\ at room-temperature (space group \textit{R$\mathit{\overline{3}}$c}).}\label{SXRDTB}

\end{table}

\begin{table}[!h]
  \caption*{Supplementary Table 2: \textbf{Magnetic symmetry} Basis vectors of the irreducible representations of the space group $Pm\bar{3}m$ for sites $1b$ with propagation vector {\textbf{k}} = (0.25 0.25 0.25).}
  \begin{tabular}[htbp]{@{}llllll@{}}
    \hline
      \multicolumn{3}{c}{Ni ($1b$)} & (1/2, 1/2, 1/2)\\
    \hline
    $\mathrm{\Gamma_2}$ & $\psi_1$ & Re  & (1, 1, 1) \\
    $\mathrm{\Gamma_3}$ & $\psi_1$ & Re  & (1, -0.5, -0.5)\\
                        &          & Im  & (0, -$\sqrt{3}$/2,  $\sqrt{3}$/2)\\
                        & $\psi_2$ & Re  & (0.5, -1,  0.5)\\
                        &          & Im  & ($\sqrt{3}$/2,  0, -$\sqrt{3}$/2) \\
    \hline
  \end{tabular}
  \label{basisvec}
\end{table}


\begin{table}[!h]
  \caption*{Supplementary Table 3: \textbf{Magnetic structure} Comparison of the measured and calculated intensities for our two models shown in Supplementary Figure~6.
	I$_{\mathrm{obs}}$: measured intensities, I$_{\mathrm{cal},\Gamma_2}$: calculated intensities for the longitudinal moment modulation, I$_{\mathrm{cal},\Gamma_3}$: calculated intensities for the helical moment arrangement. }
  \begin{tabular}[htbp]{@{}cccc@{}}
    \hline
      magnetic peak  &  I$_{\mathrm{obs}}$ &  I$_{\mathrm{cal},\Gamma_2}$ &  I$_{\mathrm{cal},\Gamma_3}$  \\
    \hline
    (0.25 0.25 0.75) &    18.3       &    26.5    &     17.4      \\
    (0.75 0.75 0.25) &    27.3       &    13.6    &     21.6      \\
    (0.75 0.75 0.75) &    22.2       &     0      &     26.6      \\
    \hline
  \end{tabular}
  \label{intensity}
\end{table}

\begin{figure}
\begin{center}
\includegraphics*[width=0.9\columnwidth]{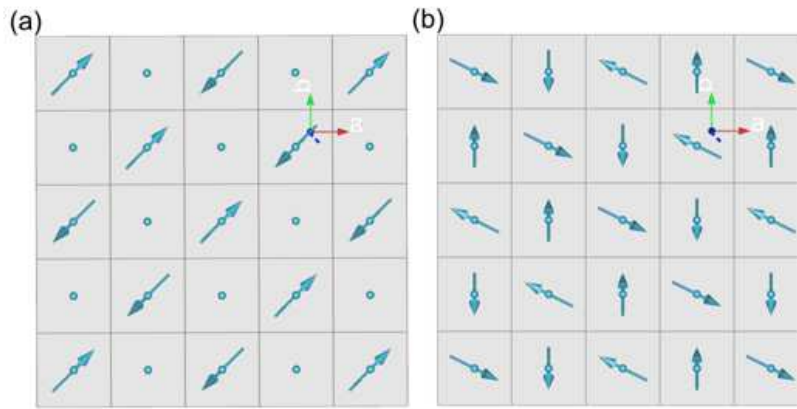}
\end{center}
\caption*{Supplementary Figure 6: \textbf{Spin structure} Spin configurations for models with (a) longitudinal modulation of the magnetic moments ($\Gamma_2$) and (b) helical arrangement of the magnetic moments perpendicular to the propagation vector  ($\Gamma_3$).}
\label{figSpin}
\end{figure}

\end{document}